\documentclass[prb,final,twocolumn,superscriptaddress]{revtex4}
\usepackage{graphicx}
\usepackage{dcolumn}
\usepackage{bm}

\begin{document}

\preprint{Preprint}

\title{Temperature dependence of crystal field excitations in CuO}
\author{Simo~Huotari}
\affiliation{Department of Physics, P.O.Box 64, FI-00014 University of
Helsinki, Finland}
\affiliation{European Synchrotron Radiation Facility, B.P.\ 220, F-38043 
Grenoble cedex, France}
\author{Laura~Simonelli}
\affiliation{European Synchrotron Radiation Facility, B.P.\ 220, F-38043 
Grenoble cedex, France}
\author{Christoph J. Sahle}
\affiliation{Department of Physics, P.O.Box 64, FI-00014 University of
Helsinki, Finland}
\author{Marco~Moretti~Sala}
\affiliation{European Synchrotron Radiation Facility, B.P.\ 220, F-38043 
Grenoble cedex, France}
\author{Roberto~Verbeni}
\affiliation{European Synchrotron Radiation Facility, B.P.\ 220, F-38043 
Grenoble cedex, France}
\author{Giulio~Monaco}
\affiliation{European Synchrotron Radiation Facility, B.P.\ 220, F-38043 
Grenoble cedex, France}
\affiliation{Physics Department, University of Trento, Via Sommarive 14, 38123 Povo (TN), Italy}
\date{\today}

\begin{abstract}
We report a study on the temperature dependence of the 
charge-neutral crystal field ($dd$)
excitations in cupric oxide,
using nonresonant inelastic x-ray scattering
(IXS) spectroscopy. 
Thanks to a very high energy resolution ($\Delta E$=60 meV),
we observe thermal effects on 
the $dd$ excitation spectrum fine structure between temperatures of 
10--320 K. With an increasing temperature,
the spectra broaden considerably.
We assign the temperature dependence of the $dd$ excitations to 
the relatively large electron-phonon coupling. 
\end{abstract}

\maketitle


Orbital physics in transition metal oxides (TMO's) remains in the
forefront of modern physics.  The complex interplay
between orbital and other electron degrees of freedom results in 
a wide range of
phenomena that pose challenges for condensed-matter physics, such as
metal-insulator transitions, superconductivity, and colossal
magnetoresistance.  \cite{tokura00}
Cupric oxide CuO (tenorite) is an antiferromagnetic insulator that
has recently attracted interest as an induced multiferroic with high 
$T_C$ (Ref.\, \onlinecite{kimura08}).
Unlike its periodic-table neighbors NiO or CoO, CuO does not have a 
cubic structure with an octahedral coordination of the metal ion,
but instead has a monoclinic unit cell and a slightly distorted
square planar local coordination. 
These CuO$_4$ units bear remarkable resemblance to
the CuO$_2$ planes in cuprate high-$T_C$ superconductors. 
The understanding of the electronic structure of CuO is thus an important
benchmark for the quest on superconductivity.
CuO also displays many intriguing properties such as 
charge-stripe ordering \cite{zheng00} and spin-phonon interaction.
\cite{chen95} 

The orbital excitations that take place within the $3d$ shell of
a transition metal ion, also called $dd$ or crystal-field excitations,
are well known probes of the local electronic structure.
They have traditionally been studied using optical 
absorption spectroscopies, \cite{ruckamp} but 
electron-energy-loss spectroscopies \cite{fromme} and resonant IXS (RIXS)  
\cite{ament11,moretti11,ghiringhelli04,ghiringhelli09,schlappa12} 
have been relatively recent introductions to the $dd$ excitation
toolbox. Even more recently, non-resonant IXS (NRIXS) has 
emerged as complementary, bulk-sensitive and high-resolution
tool for $dd$ excitations. \cite{larson07,haverkort07} Indeed,
in many transition metal oxides they have been extensively studied using 
NRIXS. \cite{larson07,hiraoka09,hiraoka11,iori12,huotari10}
A detailed theoretical framework has been formulated for the
interpretation of non-resonant IXS for studies of $dd$ excitations
via an effective operator formalism.
\cite{veenendaal08} 

Only few studies have been reported on the
crystal-field and charge-transfer excitations 
in CuO. Optical-absorption studies cannot
probe it since the band gap 
\cite{marabelli95} is smaller than the $dd$ excitation energy.
D{\"o}ring {\em et al.} \cite{doring04} studied CuO 
using RIXS by resonantly enhancing the charge-transfer excitation
at the Cu $K$ absorption edge ($E=8.98$ keV). Ghiringhelli {\em et al.}
\cite{ghiringhelli04,ghiringhelli09} have studied the $dd$ excitations
using RIXS at the Cu $L_3$ absorption edge ($E=930$ eV). They reported
$dd$ excitation spectra centered at around $\sim$ 2 eV. However,
detailed high-resolution analyses are needed in order 
to fully understand the spectral characteristics and assignment
of this fundamentally important compound.

CuO exhibits two successive magnetic transitions at $T_{N1}=213$~K 
and $T_{N2}=230$~K.\cite{forsynth88,yang89} 
Below $T_{N1}$, CuO is in an antiferromagnetic
commensurate collinear phase, and between the two
transition temperatures in an incommensurate spiral
phase. The latter phase has recently drawn interest
because it has ferroelectric properties with a very 
high ferroelectric critical temperature $T_C=T_{N2}$.\cite{kimura08}
The band gap has also been shown to have a strong dependence of
temperature, due to a relatively strong electron-phonon 
coupling.\cite{marabelli95}
The present study aimed at the determination of the $dd$ spectra
as a function of temperature, especially to see
whether the two phase transitions or electron-phonon coupling
have detectable influences on the spectral lineshape.
While for example the electronic structure of CoO has been studied 
as a function of temperature recently, \cite{kurian13,wray13}
for CuO
temperature-dependent high-energy-resolution studies have not been reported
to our knowledge.

In this article, we report high-resolution ($\Delta E=60$ meV) 
NRIXS spectra of CuO in temperatures between 10--320 K.
The observable in NRIXS is the intensity of radiation scattered
via an inelastic process where both momentum $\hbar \mathbf{q}$ 
and energy $\hbar\omega$ are transferred to the electron system.\cite{schulke07}
In the following we assume atomic units, {\em i.e.}, $\hbar=1$. 
The probability for scattering is quantified by the 
doubly differential cross section, 
which is related to the electron dynamic structure factor \cite{schulke07}  as
$$
\frac{\mathrm{d}^2\sigma}{\mathrm{d}\Omega \mathrm{d}\omega}
= \left(\frac{\mathrm{d}\sigma}{\mathrm{d}\Omega}\right)_\mathrm{Th}
S(\mathbf{q},\omega)
$$
where $(\mathrm{d}\sigma/\mathrm{d}\Omega)_\mathrm{Th}$ is the
Thomson scattering cross section, and $S(\mathbf{q},\omega)$, 
the dynamic structure factor, contains the information on the material
properties to be investigated. The same function is measured in 
electron energy loss spectroscopy (EELS). \cite{fromme} 
Both EELS and NRIXS have their advantages. In general, NRIXS has its strengths
in being bulk sensitive, yielding also access to extreme sample environments
such as high pressure, and having an access to high momentum
transfers. The $S(\mathbf{q},\omega)$ can be written as 
$$
S(\mathbf{q},\omega)= \sum_F \left| \langle F | \sum_j e^{i\mathbf{q}\cdot\mathbf{r}_j} | I \rangle \right|^2 \delta(\Omega_F-\Omega_I-\omega),
$$
where $| I \rangle$ ($\Omega_I$) and $| F \rangle$ ($\Omega_F$) are the initial and final states (energies) of the electron system, respectively, with a 
summation over all electrons $j$.
The dynamic structure factor is also related to the macroscopic dielectric function $\varepsilon(\mathbf{q},\omega)$ as
$$
S(\mathbf{q},\omega)=-\frac{n}{4\pi e^2} \mathrm{Im} [\varepsilon^{-1}(\mathbf{q},\omega)].
$$
This equivalence is often used to relate optical spectra and dielectric screening to the
results of an energy-loss experiment such as EELS or NRIXS.\cite{huotari10,weissker06}
The theoretical framework on how NRIXS can access dipole-forbidden excitations
in different systems has been laid down in, {\em e.g.,} Refs.\, \onlinecite{soininen05,haverkort07,caciuffo10,veenendaal08,gordon08}.

CuO has a monoclinic crystal structure (space group
$C2/c$) (Ref.\, \onlinecite{asbrink}) 
with four nonequivalent Cu and O sites in the primitive
unit cell.\cite{forsynth88,yang89} The lattice parameters are $a=4.68$ \AA, $b=3.42$ \AA,
$c=5.13$ \AA, $\alpha=\gamma=90^\circ$, $\beta=99.5^\circ$. 
The structure of CuO can be thought  
to consist of two different kinds of CuO$_4$ plaquettes that are 
at an angle of 77.84$^\circ$ with respect to each other. 
The orientation
of the $\mathbf{q}$-vector with respect to the planes is thus
in general an average over the two nonequivalent planes.

In the following discussion we assume Cu$^{2+}$ ions in a CuO$_4$ 
plaquette with a $D_{4h}$ point group symmetry. Within the crystal
field model, \cite{figgis00} the local field 
splits the $3d$ energy levels to $a_{1g}$ ($d_{z^2}$),
$b_{1g}$ ($d_{x^2-y^2}$), $b_{2g}$ ($d_{xy}$), and doubly degenerate $e_g$ ($d_{xz}$ and $d_{yz}$). 
In the ground state, the hole occupies the $d_{x^2-y^2}$ orbital. More refined calculations \cite{eskes90,huang11,lany13,iori12,rodl08,rodl09,rodl12,schron12,takahashi97,haverkort12}
can be done in order to include Cu-O hybridization and band structure
but the crystal field model is sufficient to capture the overall energy-level picture.
The CuO$_4$ plaquettes in CuO are not square, but rather nerly
rectangular
parallelograms with side lenghts of 2.62 \AA~and 2.90 \AA, 
and exhibit two different Cu-O distances (1.95 \AA~and 1.96 \AA). 
This lifts the degeneracy of the $xz$ and $yz$
orbitals, with an energy splitting that is expected to be $\sim$60 meV
(Ref.\, \onlinecite{huang11}). This splitting 
would be expected to be resolvable with the energy
resolution of the current study. 

The experiment was performed at the beamline ID16 of the European 
Synchrotron Radiation Facility. \cite{id16}
The incident photon beam
was monochromatised using a combination of Si(111)
premonochromator and a Si(444) channel-cut to a
bandwidth of 40~meV. 
The beam was focused using a toroidal mirror
to a spot size of 30$\times$100~$\mu$m$^2$ (V$\times$H) on the sample. 
We used a spectrometer designed for NRIXS experiments.\cite{verbeni09} 
It was equipped with six diced Si($444$) analysers
fixed at a Bragg angle of 88.5$^\circ$, with a photon energy of 7.9~keV.
The total energy resolution was 60~meV.  
The sample temperature was controlled using a miniature He-flow cryostat.
The sample was a single crystal of CuO (the same as used in
Ref.\ \onlinecite{doring04}).
The spectra were measured at a fixed momentum transfer 
value $q=|\mathbf{q}|=(7.5\pm 0.1)$ \AA\, 
with the average $\mathbf{q}$ in the direction 
$\mathbf{q}~||~[540]$.
Based on the expected angular dependence of $dd$ excitations
\cite{veenendaal08}, 
in this geometry the $xz/yz$ and $z^2$ peaks are expected to be 
excited most strongly, with the $xy$ peak to be weak.
The spectra were measured in several temperatures 
between 10 K and 320 K.

\begin{figure}
\includegraphics[width=.9\linewidth]{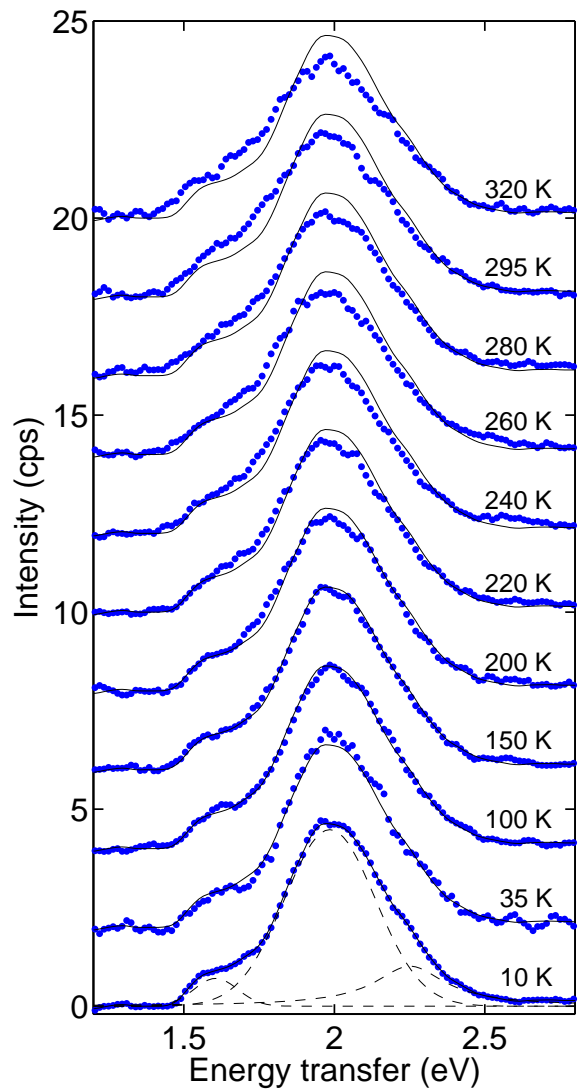}
\caption{\label{fig:spectra}(Color online) CuO d-d excitation spectra
(dots) measured as a function of temperature. A smoothed version of the 
curve measured at $T=10$ K (line) is shown for reference throughout.
The dashed lines drawn together with the spectrum at $T=10$~K are the
result of fitting three Pearson VII functions to that spectrum.}
\end{figure}

All collected spectra as a function of temperature, after subtracting 
a sloping background due to the quasielastic line tail, are shown in 
Figure \ref{fig:spectra}. The spectra have been normalised to 
have the same area  between 1--3 eV.
The spectra can be broken into a few components: a main peak at 2 eV
and a weaker peak manifesting itself as a shoulder at 1.6 eV, and 
an even weaker shoulder (mainly visible at lowest temperatures) at 2.2 eV.
A recent {\em ab initio} calculation\cite{huang11} predicts
the peaks to be assigned as, 
from lowest to highest energy, excitations from the $x^2-y^2$ orbital to 
the  $xy$, $xz/yz$, and $z^2$ orbitals.
The assignment of the peaks to specific $d$-orbital can 
be confirmed with either angular dependence \cite{moretti11}
or their dependence on 
$q$ (Ref.\, \onlinecite{veenendaal08}), which 
are subject of consequent studies.
In this particular work, 
we concentrate on the overall temperature dependence of the 
spectra.

The main effect of increasing temperature from $T=10$ K 
is a clear broadening of the overall spectral shape.
Due to the broadening, 
the low-energy shoulder seems to merge into the
main peak and is nearly undetectable at room temperature. 
Thus, an important result is the bandwidth of the excitations: with a 60-meV
energy resolution the main peak has a width of the order of 400 meV
even at $T=10$ K. This is partly due to the overlap of the 
$xz/yz$ and the $z^2$ excitation but even then the individual components
have a width of about $\sim$300--400 eV.
While in the orbital ionic picture the $3d$ states are expected to have a 
very narrow line shape, when switching on the band structure 
the $3d$ states gain 
non-negligible bandwidth due to the electron-ion interaction and hybridization.
\cite{ching89,wu06,lany13,heinemann13}
The observed width extrapolated to $T=0$~K
thus may reflect the width of the density of states of the occupied and
unoccupied $3d$ bands.\cite{iori12} 
Time-dependent density functional theory that takes into account
band structure, realistic transition matrix elements and
local field effects could possibly explain the spectral linewidth and shape
in a more detailed way. \cite{gurtubay05,huotari09,v2o3} Another way
of viewing this is to consider 
that $dd$ excitations couple to the continuum states 
beyond the band gap. Thus the
temperature dependence could be quantified by relating the $dd$ width 
as a function of temperature to the thermal behavior of the band gap itself.
It should be noted that the $dd$ excitations in NiO also have 
non negligible band width of the order of $200$ meV,
\cite{huotari-nio-nrixs} even though the
band gap  is larger in NiO ($\sim$4 eV in comparison to $E_g\sim$1.35 eV of
CuO in room temperature).

\begin{figure}
\includegraphics[width=\linewidth]{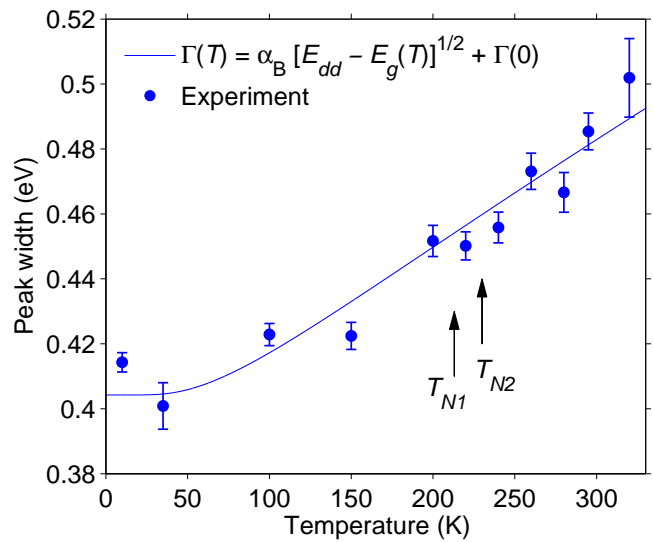}
\caption{\label{fig:cuopeakwidth} 
The $dd$ peak width (points) compared to a fit based on the 
band gap width $E_g$ as a function of temperature (line).}
\end{figure}

In order to quantify the change in the shape  as a function of temperature,
we fitted the spectra by Pearson VII functions. \cite{pearson7}
An example of such fit in the case of $T=10$~K
is shown in Figure \ref{fig:spectra}.
Since the 
lowest and highest energy peaks are weak in this geometry, their position
nor width can not be fitted
very reliably. However, the determination of the width of the main peak 
at 2.0 eV can be done with a very high 
accuracy.
The resulting fitted values for the 
width (full-width-at-half maximum, FWHM)
of the 2.0-eV peak are shown in Fig.\ \ref{fig:cuopeakwidth} 
as a function of temperature.
One important result is that
the peak width across different temperatures
does not have a significant relation to the magnetic transitions as it does not
exhibit significant changes across either transition temperature.
Instead,
the thermal behavior of the peak width seems to be rather smooth
across the studied temperature range.

Thus, the effect of temperature on $dd$ excitations in CuO is clear
and easily detectable.
The width of the band gap of CuO has been reported to vary from 
$E_g=1.55$ eV at $T=0$ K
to 1.35 eV at $T=300$ K.\cite{marabelli95} If the relatively large 
width of the $dd$ excitations
($\sim$ 400 meV) is due to interaction with continuum states 
owing to the presence of the band gap,
the decreasing gap width with increasing temperature could explain the 
observed behavior. In this scenario, an increase of the density of states
at, or near to,
the energy of $dd$ excitations could increase the width of the 
$dd$ peaks. The temperature-dependence of the optical gap has been explained
to be due to the large electron-phonon coupling. \cite{marabelli95}
Electron-phonon coupling thus seems a natural reason
for the temperature-dependence of the $dd$ excitations as well.

Using a Bose-Einstein
statistical factor for phonons with average energy of $k_B\theta$,
the gap energy as a function of temperature
\cite{marabelli95} can be fitted to a form\cite{lautenschlager}
$$
E_\mathrm{g}(T) = E_\mathrm{B}-a_\mathrm{B} \left[ 1+ \frac{2}{e^{\theta/T}-1} \right],
$$
with $E_\mathrm{B}=1.66$ eV, $a_\mathrm{B}=0.1$ eV and $\theta=196$ K.
We assume a density of states above the gap of the free-electron form
$\rho(E) \propto \sqrt{E-E_g}$, when $E \geq E_g$, and $\rho(E)=0$ when $E<E_g$.
Assuming a linear dependence of the 
$dd$ spectral linewidth $\Gamma$ on the density of states at the $dd$ excitation energy,
$$
\Gamma(T) =  \alpha_\mathrm{B} \sqrt{E_{dd}-E_g(T)} + \Gamma(0),
$$
we get a good agreement with the experiment with
$\Gamma(0) = 0.0569$ eV and $\alpha_\mathrm{B}=0.526$ eV$^{1/2}$. The resulting fit is
shown in Fig.~\ref{fig:cuopeakwidth}. Even if the observed peak here is a
superposition of different $dd$ excitations, which gives 
a non-negligible contribution to $\Gamma(0)$, the temperature dependence
is the most interesting result here. 
The fit agreement is good, yielding insight that the interaction with 
the continuum states could be the underlying reason for the $dd$ excitation
lineshape. The temperature dependence shows a good agreement with a
correspondence to the band gap variation
that in turn has its underlying reasons in the electron-phonon coupling.

The {\em ab initio} optical absorption spectrum in the
range of $dd$ excitations in NiO has been calculated based on 
molecular dynamics simulations in finite temperature,
\cite{domingo12} but 
to our knowledge, such calculations do not exist for CuO. 
A finite distribution of
Cu-O bond lengths in finite temperatures, due to thermal disorder, 
is expected to have an 
effect similar to the one observed here. This is because the 
$dd$ excitation energy is proportional to $a_\mathrm{Cu-O}^{-5}$, where 
$a_\mathrm{Cu-O}$ is the Cu-O bond distance. 
Further, in principle, the coupling to the lattice 
could be possibly quantified from phonon
parameters. \cite{kuzmenko,homes95} However, the good agreement obtained
by using the phenomenological fit to the band-gap energy, 
already gives important insights to the coupling of the
$dd$ excitations to the band gap and the electron-phonon coupling.

In conclusion,
we have measured the $dd$ excitation spectra of bulk CuO with
non-resonant IXS
with high energy resolution as
a function of temperature. 
Most importantly, the study reveals 
the coupling of the orbital excitations to phonons
via the temperature dependence of the spectral shape. 
The spectral changes can be understood of being due to the
interaction with the continuum states above the band gap, which
in turn depends on temperature due to electron-phonon coupling.
These results are important to understand 
the bandwidth related to the $dd$ excitations and their temperature
dependence.

\begin{acknowledgments}
Beamtime was granted by the European Synchrotron Radiation Facility.
Funding was provided by 
the Academy of Finland (Grants 1256211, 1127462, and 1259526)
and University of Helsinki Research Funds (Grant 490076). 
We are grateful for C.~Henriquet, M.-C.~Lagier, 
and the whole beamline ID16 team and support groups 
for expert assistance, advice, and encouragement in the experiment. 
We would like to thank M.~W.~Haverkort, M.~Hakala, M.~Gatti,
and C.~R{\"o}dl for fruitful discussions.
\end{acknowledgments}

\end{document}